\documentclass[twocolumn,showpacs,preprintnumbers,amsmath,amssymb]{revtex4}
\usepackage{graphicx,color}% Include figure files
\usepackage{bm}% bold math

\begin{document}

%%%%%%%%%%%%%%%%%%%%%%%%%%%%%%%%%%%%%%%%%%%%%%%%%%%%%%%%%%%%%%%%%%%%%%%%%%
\title{Dihadron Tomography of High-Energy Nuclear Collisions \\
in Next-to-Leading Order perturbative QCD}
%%%%%%%%%%%%%%%%%%%%%%%%%%%%%%%%%%%%%%%%%%%%%%%%%%%%%%%%%%%%%%%%%%%%%%%%%%

\author{Hanzhong Zhang$^a$, Joseph~F. Owens$^b$, Enke Wang$^a$ and Xin-Nian Wang$^c$}

\affiliation{$^a$Institute of Particle Physics, Huazhong Normal University,
         Wuhan 430079, China\\
$^b$Physics Department, Florida State University, Tallahassee,
          FL 32306-4350, USA\\
$^c$Nuclear Science Division, Lawrence Berkeley Laboratory,
         Berkeley, California 94720, USA}

%\vspace{-1.5in}
%{\hfill LBNL-62151}
%\vspace{1.4in}

\preprint{LBNL-62151}

\begin{abstract}

Dihadron spectra in high-energy heavy-ion collisions are studied 
within the next-to-leading order perturbative QCD parton model with modified 
jet fragmentation functions due to jet quenching. High-$p_T$ back-to-back
dihadrons are found to originate mainly from jet pairs produced close and
tangential to the surface of the dense matter. However, a substantial
fraction also comes from jets produced at the center with finite energy
loss. Consequently, high-$p_T$ dihadron spectra are found
to be more sensitive to the initial gluon density than the single hadron
spectra that are more dominated by surface emission. A simultaneous $\chi^2$-fit 
to both the single and dihadron spectra can be achieved within a
range of the energy loss parameter $\epsilon_0=1.6-2.1$ GeV/fm.
Because of the flattening of the initial jet production 
spectra at $\sqrt{s}=5.5$ TeV, high $p_T$ dihadrons
are found to be more robust as probes of the dense medium.

\vspace{12pt}

%\noindent {\em PACS numbers:} 12.38.Mh, 24.85.+p; 25.75.-q
\end{abstract}


\maketitle

Jet quenching as discovered in high-energy heavy-ion collisions at
the Relativistic Heavy-ion Collider (RHIC) is manifested in both
the suppression of single inclusive hadron spectra at large transverse
momentum ($p_T$) \cite{phenix} and the disappearance of the typical
back-to-back jet structure in dihadron correlation in vacuum \cite{star}.
Since jet quenching is caused by parton energy loss which in turn depends
on the gluon density and transport coefficient of the medium, detailed
study of the suppression of large $p_T$ hadron spectra and correlations
can shed light on the properties of the dense medium \cite{review} .

In heavy-ion collisions, the spatial distribution
of the initial jet production points is given by the nuclear overlap,
$T_{AB}({\bf b},{\bf r})=t_A(r)t_B(|{\bf b}-{\bf r}|)$ with $t_A(r)$
being the thickness function of each nucleus. The suppression factor
of the leading hadrons from jet fragmentation will
depend on the total parton energy loss which in turn is related to
the weighted gluon density integrated along the jet propagation path.
Therefore, measurements of
large $p_T$ hadron suppression can be directly related to the averaged
gluon density and medium transport coefficient. As the average gluon
density increases with colliding energy and centrality, one should
expect continued decrease of the suppression factor until 
particle production in the outer corona of the
medium dominates the single hadron spectra \cite{eskola}. 
In this case, the suppression
factor of high-$p_T$ hadron spectra will lose its effectiveness as
a sensitive probe since it only depends on the thickness of
the outer corona which varies very slowly with the initial gluon
density. In this Letter we will employ for the first time a
next-to-leading order (NLO) perturbative QCD (pQCD) parton model to
study the suppression of both single and  dihadron spectra due to jet
quenching. In particular, we will investigate the robustness of
back-to-back dihadron spectra as a probe of the initial gluon density
when single hadron spectra suppression become fragile.

Within the NLO pQCD parton model, large $p_T$ hadron
production cross section in $p+p$ collisions can be expressed as a
convolution of NLO parton-parton scattering cross sections, parton
distributions inside nucleons and parton fragmentation
functions (FF). The calculations discussed in this Letter are carried
out within a NLO Monte Carlo based program \cite{owens} which
utilizes two-cutoff parameters, $\delta_s$ and $\delta_c$,
to isolate the soft and collinear divergences in the squared
matrix elements of the $2\rightarrow 3$ processes. The regions containing
the divergences are integrated over in n-dimensions and the results are
combined with the squared matrix elements for the $2\rightarrow 2$
processes. This results in a set of two-body and three-body weights,
each of which depends on the cut-offs. However, this dependence cancels
when the weights are combined in the calculation of physical observables.

For the study of large $p_T$ single and dihadron production in $A+A$
collisions, we assume that the initial hard scattering cross sections are
factorized as in $p+p$ collisions. As in the lowest-order (LO) pQCD parton model
study \cite{Wang:2003mm}, we further assume that the effect of final-state
interaction between produced parton and the bulk medium can be described
by the effective medium-modified FF's,
\begin{eqnarray}
D_{h/c}(z_c,\Delta E_c,\mu^2) &=&(1-e^{-\langle \frac{L}
{\lambda}\rangle}) \left[ \frac{z_c^\prime}{z_c}
D^0_{h/c}(z_c^\prime,\mu^2) \right.
 \nonumber \\
& &\hspace{-0.9in} \left. + \langle \frac{L}{\lambda}\rangle
\frac{z_g^\prime}{z_c} D^0_{h/g}(z_g^\prime,\mu^2)\right] +
e^{-\langle\frac{L}{\lambda}\rangle} D^0_{h/c}(z_c,\mu^2),
\label{eq:modfrag}
\end{eqnarray}
where $z_c^\prime=p_T/(p_{Tc}-\Delta E_c)$,
$z_g^\prime=\langle L/\lambda\rangle p_T/\Delta E_c$ are the rescaled
momentum fractions, $\Delta E_c$ is the average radiative parton
energy loss and $\langle L/\lambda\rangle$ is the number of
scatterings. The FF's in vacuum $D^0_{h/c}(z_c,\mu^2)$ are given by the
KKP parameterization \cite{KKP}.

According to recent theoretical studies \cite{gvw,ww,sw02} the
total parton energy loss in a finite and expanding medium can be
approximated as a path integral,
\begin{equation}
\Delta E \approx \langle \frac{dE}{dL}\rangle_{1d}
\int_{\tau_0}^{\infty} d\tau \frac{\tau-\tau_0}{\tau_0\rho_0}
\rho_g(\tau,{\bf b},{\bf r}+{\bf n}\tau),
\end{equation}
for a parton produced at a transverse position ${\bf r}$ and 
traveling along the direction ${\bf n}$.
$\langle dE/dL\rangle_{1d}$ is the average parton energy loss per
unit length in a 1-d expanding medium with an initial uniform gluon
density $\rho_0$ at a formation time $\tau_0$ for the medium gluons. 
The energy dependence of the energy loss is
parameterized as
\begin{equation}
 \langle\frac{dE}{dL}\rangle_{1d}=\epsilon_0 (E/\mu_0-1.6)^{1.2}
 /(7.5+E/\mu_0),
\label{eq:loss}
\end{equation}
from the numerical results in Ref.~\cite{ww} in which thermal
gluon absorption is also taken into account.
%in the calculation of parton energy loss.
The parameter $\epsilon_0$ should be proportional
to
%the initial gluon density
$\rho_0$.

Neglecting transverse expansion, the gluon density distribution
in a 1-d expanding medium in $A+A$ collisions at impact-parameter
${\bf b}$ is assumed to be proportional to the transverse profile of
participant nucleons ,
\begin{equation}
\rho_g(\tau,{\bf b},{\bf r})=\frac{\tau_0\rho_0}{\tau}
        \frac{\pi R^2_A}{2A}[t_A({\bf r})+t_A(|{\bf b}-{\bf r}|)].
\end{equation}
The average number of scatterings along the parton propagating
path is
\begin{equation}
\langle L/\lambda\rangle =\int_{\tau_0}^{\infty}
\frac{d\tau}{\rho_0\lambda_0} \rho_g(\tau,{\bf b},{\bf r}+{\bf n}\tau).
\end{equation}
Here, we neglect the time-dependence of the cross section
and characterize it by the mean free path $\lambda_0$ via
$\sigma_0=1/(\rho_0\lambda_0)$ and a hard-sphere nuclear
distribution is used. The parton
distributions per nucleon $f_{a/A}(x_a,{\bf r})$ inside the
nucleus are assumed to be factorizable into the parton
distributions in a nucleon given by CTEQ6M parameterization
\cite{CTEQ} and the new HIJING parameterization \cite{lw02} of the
impact-parameter dependent nuclear modification factor, including
the isospin dependence.

\begin{figure}
\begin{center}
\includegraphics[width=85mm]{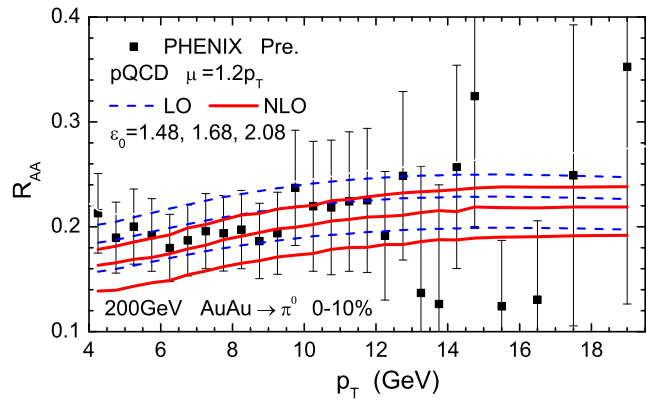}
\end{center}
\vspace{-12pt}
\caption{\label{fig:rau0-10}
 (color online). Nuclear modification factors for $\pi^0$ spectra in LO (dashed)
and NLO pQCD (solid) in $Au+Au(0-10\%)$ collisions at $\sqrt{s}=200$ GeV
with different values of energy loss parameter $\epsilon_0=1.48$, 1.68
and 2.08 GeV/fm (from top to bottom) compared with data \protect\cite{akiba}.}
\end{figure}

The calculated single inclusive $\pi^0$ and $\pi^+ + \pi^-$ spectra
in the NLO pQCD parton model in $p+p$ collisions agree well with the
experimental data at the RHIC energy with the factorization scale
in the range $\mu=0.9 \sim 1.5 p_T$ \cite{zoww}. We use the same
factorization scale in both $p+p$ and $A+A$ collisions in our
calculation. Shown in Fig.~\ref{fig:rau0-10} are the nuclear
modification factors,
\begin{eqnarray}
R_{AA}=\frac{d\sigma_{AA}/dp_T^2dy}
{\int d^2b\, T_{AA}({\bf b})d\sigma_{NN}/dp_T^2dy}\,, \label{eqn:modifactoer}
\end{eqnarray}
for single inclusive $\pi^0$ spectra calculated in both leading-order
(LO) and NLO pQCD parton model as compared to the PHENIX data
on central $Au+Au$ collisions at $\sqrt{s}=200$ GeV. The factorization
scale in the NLO result is set $\mu=1.2 p_T$ though the nuclear
modification factor $R_{AA}$ is not at all sensitive to the choice of $\mu$.
However, $R_{AA}$ in NLO calculation is always smaller than the LO
result because of the relative larger ratio of gluon/quark
jets in NLO than in LO calculation and gluon energy loss is assumed to be
9/4 larger than that of a quark. In both calculations, we have chosen
the parameters as $\mu_0=1.5$ GeV, $\epsilon_0\lambda_0=0.5$ GeV
and $\tau_0=0.2$ fm/$c$. 
The results are not sensitive to small values of $\tau_0$.

%$\epsilon_0=1.68$ GeV/fm,
%and $\lambda_0=1/\sigma\rho_0=0.3$ fm.

%\vspace{-6pt}
%%%%%%%%%%%%%%%%%%%%%%%%%%%%%%%%%%%%%%%%%%%%%%%%%%%%%%%%%%%%%%%%%%%%%%%%%
\begin{figure}
\begin{center}
\includegraphics[width=85mm]{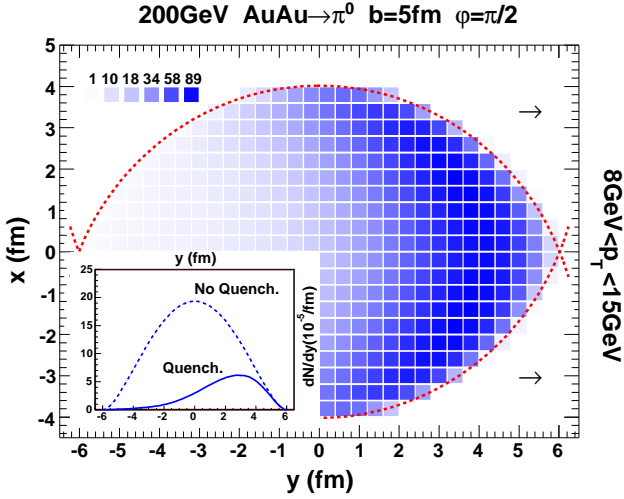}
\end{center}
\vspace{-12pt}
 \caption{\label{fig:con-sin-y0}
(color online). Spatial transverse distribution (arbitrary normalization) of the initial
parton production points that contribute to the final $\pi^0$ at $8 < p_T < 15$
GeV along $\phi=\pi/2$. The insert is the same distribution projected onto
the $y$-axis.}
\end{figure}

\begin{figure}
\begin{center}
\includegraphics[width=85mm]{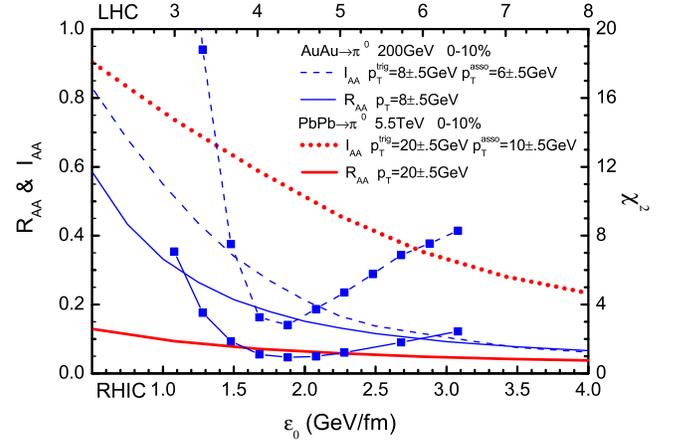}
\end{center}
\vspace{-12pt}
\caption{\label{fig:eps-dep}
(color online). 
The suppression factors for single ($R_{AA}$),  dihadron ($I_{AA}$) spectra
at fixed transverse momentum and $\chi^2 /d.o.f.$ (curves with
filled squares) in fitting experimental data on single \protect\cite{akiba}
($p_T=4 - 20 $ GeV/$c$) and away-side spectra \protect\cite{Adams:2006yt}
($p_T^{\rm trig}=8-15$ GeV, $z_T=0.45-0.95$) in central $Au+Au$ collisions 
at $\sqrt{s}=200$ GeV as functions of the initial energy loss 
parameter $\epsilon_0$.} 
\end{figure}

Because of jet quenching, the dominant contribution to
the measured single hadron spectra at large $p_T$ comes from those
jets that are initially produced in the outer corona of the
overlap region toward the direction of the detected hadron.
This is clearly illustrated in Fig.~\ref{fig:con-sin-y0} by the spatial
distribution of the production points of those jets that have survived
the interaction with the medium and contribute to the
measured spectra. Jets produced in the region away from the detected
hadron are severely suppressed due to their large energy
loss and don't contribute much to the final hadron spectra. As pointed
out in Ref.~\cite{eskola}, when jets produced in the inner part of the
overlapped region are completely suppressed due to large initial gluon
density, the final large $p_T$ hadron production is dominated
by ``surface emission''. High-$p_T$ hadron yield via such
surface emission should be proportional to the thickness of the
outer corona which decreases with the initial gluon density.
Therefore, the suppression factor for single hadron spectra should
continue to decrease with the initial gluon density
as shown in Fig.~\ref{fig:eps-dep}. The variation, nevertheless,
is very weak when surface emission becomes dominant and single
hadron suppression is no longer a sensitive probe of the initial
gluon density.

%%%%%%%%%%%%%%%%%%%%%%%%%%%%%%%%%%%%%%%%%%%%%%%%%%%%%%%%%%%%%%%%%%%%%%%%%%%

To find robust probes of the high initial gluon
density when single hadron spectra suppression becomes fragile, we will
study back-to-back dihadron spectra within NLO pQCD parton model in this
Letter. To quantify dihadron spectra, a hadron-triggered
fragmentation function,
\begin{equation}
D_{AA}(z_T,p_T^{\rm trig}) \equiv p_T^{\rm trig}\frac{
  d\sigma^{h_1h_2}_{AA}/dy^{\rm trig}
dp^{\rm trig}_T dy^{\rm asso}dp_T^{\rm asso}}
  {d\sigma^{h_1}_{AA}/dy^{\rm trig}dp^{\rm trig}_T}\,,
%\label{Daa-tmp}
\end{equation}
was introduced \cite{Wang:2003mm} as a function
of $z_T=p^{\rm asso}_T/p^{\rm trig}_T$, which is
essentially the away-side hadron spectrum associated with a triggered
hadron. Both hadrons are limited to the central rapidity region
$|y^{\rm trig, asso}|<0.5$ and the azimuthal angle relative to the
triggered hadron is integrated over $|\Delta\phi|>2.5$.

\begin{figure}
\begin{center}
\includegraphics[width=85mm]{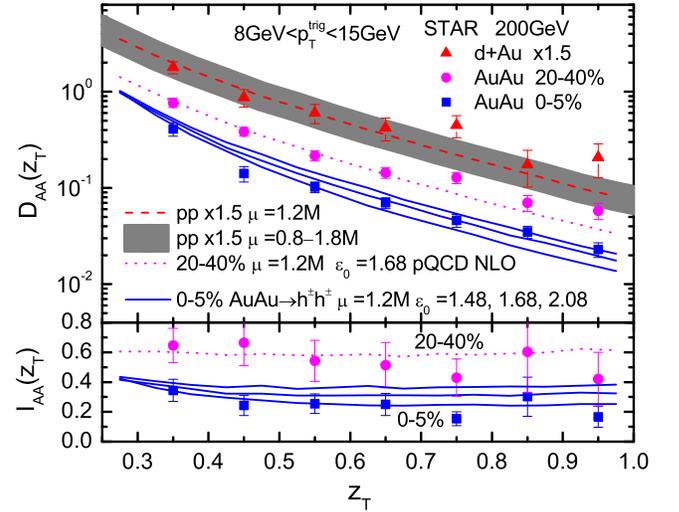}
\end{center}
\vspace{-12pt}
\caption{(color online). Hadron-triggered fragmentation functions $D_{AA}(z_T)$
and the medium modification factors $I_{AA}(z_T)$ in NLO pQCD as compared to
the data \protect\cite{Adams:2006yt}.}
\label{fig:Daa}
\end{figure}

As in the NLO calculation of single hadron spectra, one also has
to fix the factorization scale in the NLO calculation of dihadron
spectra, which is chosen to be the invariant mass of the
dihadron $M^2=(p_1+p_2)^2$. The NLO results on associated hadron
spectra $D_{pp}(z_T,p_T^{\rm trig})$ in $p+p$ collisions are compared
to the $d+Au$ data at $\sqrt{s}=200$ GeV in Fig.~\ref{fig:Daa}, assuming
no nuclear effects in $d+Au$ collisions. The overall normalization of
NLO results is quite sensitive to the factorization scale as indicated
by the shaded region corresponding to $\mu=0.8 - 1.8 M$. We will
use $\mu=1.2 M$ in this study with which NLO results describe the
experimental data well. We use the same scale in the calculation of
dihadron spectra in $Au+Au$ collisions and the results
for $D_{AA}(z_T,p_T^{\rm trig})$ agree well with STAR
data as shown in Fig.~\ref{fig:Daa} using the same energy loss
parameter $\epsilon_0=1.68-2.08$ GeV/fm. The centrality dependence of
the data is also reproduced by the NLO calculation \cite{zoww}.
The nuclear modification factor of the hadron-triggered fragmentation
function
\begin{equation}
I_{AA}=\frac{D_{AA}(z_T,p_T^{\rm trig})}{D_{pp}(z_T,p_T^{\rm trig})}
\end{equation}
as plotted in the lower panel is coincidentally similar to
the modification factor for single hadron spectra $R_{AA}$.

%\vspace{-6pt}

%\vspace{-12pt}
%%%%%%%%%%%%%%%%%%%%%%%%%%%%%%%%%%%%%%%%%%%%%%%%%%%%%%%%%%%%%%%%%%%%%%%%%%%

We have adjusted the energy loss parameter $\epsilon_0$ to fit
experimental data on both single ($d\sigma_{AA}/dp_Tdy$
between $p_T=4 - 20$ GeV/$c$) and away-side spectra [$D_{AA}(z_T)$]
in the most central $Au+Au$ collisions at $\sqrt{s}=200$ GeV. The best
fits occur for $\epsilon_0=1.6 - 2.1$ GeV/fm as shown by the
$\chi^2$ in Fig.~\ref{fig:eps-dep}. The fact that both $\chi^2$'s
reach their minima in the same region for two different measurements
is highly nontrivial, providing convincing evidence for the
jet quenching description.

\begin{figure}
\begin{center}
\includegraphics[width=85mm]{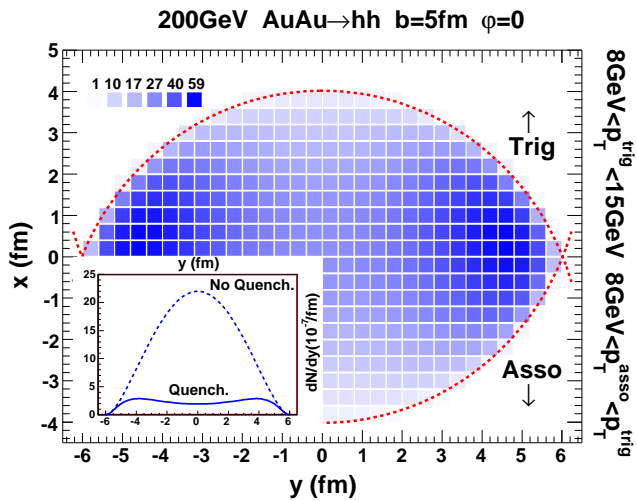}
\end{center}
\vspace{-12pt}
 \caption{\label{fig:di-x0}
 (color online). The same as Fig.~\ref{fig:con-sin-y0} except for dihadrons
 along the direction $\phi=0$ and $\pi$.}
\end{figure}

Because of trigger bias, most of
the contribution to dihadron spectra comes from dijets produced
close and tangential to the surface of the overlapped region,
as shown in Fig.~\ref{fig:di-x0}. However, there are still
about 25\% of the contribution coming from dijets near the
center of the overlapped region. These jets truly punch
through the medium and emerge after finite amount of energy loss.
This is why dihadron spectra is slightly more sensitive
to $\epsilon_0$ than the single spectra as was noted in
Refs.~\cite{dainese,renk}.
As one further increases the initial gluon density, the fraction of
these punch-through jets will also diminish and the final dihadron
spectra will be dominated by the tangential jets in the outer
corona. Dihadron spectra at RHIC will also lose its sensitivity to
the initial gluon density of the medium as shown in Fig.~\ref{fig:eps-dep}
Even if one includes the effect of transverse expansion of
the bulk matter, the above results remain qualitatively the
same \cite{renk}.

%%%%%%%%%%%%%%%%%%%%%%%%%%%%%%%%%%%%%%%%%%%%%%%%%%%%%%%%%%%%%%%%%%%%%%%%%%%

Also shown in Fig.~\ref{fig:eps-dep} are the single and dihadron
suppression factors at the LHC energy as a function of $\epsilon_0$
for fixed values of $p_T^{\rm trig}=20$ GeV and $p_T^{\rm asso}=10$ GeV.
Because of the flattening of the overall spectra shape at the LHC energy,
back-to-back dihadron spectra at a given $p_T$ have more contribution
from dijets with higher initial energy. They are therefore less suppressed
than at the RHIC energy and are more sensitive to the initial gluon density.
The single hadron spectra on the other hand are increasingly dominated by
surface emission and the suppression factor $R_{AA}$ becomes much smaller
than $I_{AA}$ for dihadrons.
%For a hypothetically large initial gluon
%density or $\epsilon_0$, surface emission would again dominate dihadron
%production and $I_{AA}$ and $R_{AA}$ would become similar.
From a model estimate of the bulk hadron production at LHC \cite{lw02}, the
energy loss parameter is about $\epsilon_0\approx 5$ GeV/fm in central $Pb+Pb$
collisions. The dihadron suppression factor $I_{AA}$ at around
this value of $\epsilon_0$ is significantly
larger than the single hadron suppression factor $R_{AA}$ and is
more sensitive to the initial gluon density.

%It is worthwhile to mention that the original prediction
%for $I_{AA}$ \cite{xn04} in LO pQCD model was made with the
%lower value of the energy loss parameter $\epsilon_0$
%extracted from fitting the single hadron suppression to the
%experimental data. That is why the predicted modification
%factor for the hadron-triggered fragmentation factor is about
%25\% higher than the data.

In summary, we have used the NLO pQCD parton model with effective
modified FF's due to radiative parton energy loss to study both 
single and dihadron spectra in high-energy heavy-ion collisions. 
We found that the surface emission dominates the single
hadron production process within the range of initial gluon density
in the most central $Au+Au$ collisions at RHIC. However, there is
still a significant fraction of dijets that are produced in the center
of the dense medium and contribute to the final dihadron spectra after
losing finite amount of energy. Therefore, dihadron spectra are more
robust probes of the initial gluon density in the most central
$Au+Au$ collisions at RHIC while single hadron spectra become less
sensitive. A simultaneous $\chi^2$-fit 
to both the single and dihadron spectra can be achieved within a
narrow range of the energy loss parameters $\epsilon_0=1.6-2.1$ GeV/fm.
If the initial gluon density at RHIC were to increase further,
dihadron spectra at a fixed $p_T$ would also be dominated by the surface
emission and become insensitive to the initial gluon density.
At LHC, however, the flattening of the initial jet production spectra
leads to an increase in the dihadron suppression factor. Dihadrons will
therefore remain more robust probes within a large range of initial
gluon density.

We thank P. Jacobs for helpful discussions. This work was supported 
by DOE under contracts No. DE-AC02-05CH11231 and
No. DE-FG02-97IR40122, by NSFC under Project No. 10440420018, No. 10475031
and No. 10635020, and by MOE of China under projects No. NCET-04-0744,
No. SRFDP-20040511005 and No. IRT0624.

\end{document}